\documentclass[aps,prb,twocolumn,showpacs,superscriptaddress,floatfix]{revtex4-1}

\pdfoutput=1

\usepackage{amsmath,amssymb}
\usepackage{graphicx}
\usepackage{color}
\usepackage{rotating}
\usepackage{bm}
\usepackage{setspace}
\usepackage{hyperref}
\usepackage{float}
\usepackage{soul}
\usepackage{color}

\hypersetup{
colorlinks=true,
urlcolor= blue,
citecolor=blue,
linkcolor= blue,
bookmarks=true,
bookmarksopen=false,
}

\renewcommand{\vec}[1]{\bm{#1}}

\begin{document}

\preprint{APS/123-QED}

\title{Electronic Structure Theory of Weakly Interacting Bilayers}

\author{Shiang Fang}
\affiliation{Department of Physics, Harvard University, Cambridge, Massachusetts 02138, USA.}
\author{Efthimios Kaxiras}
\affiliation{Department of Physics, Harvard University, Cambridge, Massachusetts 02138, USA.}
\affiliation{John A. Paulson School of Engineering and Applied Sciences, Harvard University, Cambridge, Massachusetts 02138, USA.}

\date{\today}

\begin{abstract}
We derive electronic structure models for weakly interacting bilayers such as graphene-graphene and graphene-hexagonal boron nitride, based on density functional theory calculations followed by Wannier transformation of electronic states. These transferable interlayer coupling models can be applied to investigate the physics of bilayers with arbitrary translations and twists. The functional form, in addition to the dependence on the distance, includes the angular dependence that results from higher angular momentum components in the Wannier $p_z$ orbitals. We demonstrate the capabilities of the method by applying it to a rotated graphene bilayer, which produces the analytically predicted renormalization of the Fermi velocity, van Hove singularities in the density of states, and Moir\'{e} pattern of the electronic localization at small twist angles. We further extend the theory to obtain the effective couplings by integrating out neighboring layers. This approach is instrumental for the design of van der Walls heterostructures with desirable electronic features and transport properties and for the derivation of low-energy theories for graphene stacks, including proximity effects from other layers.
\end{abstract}

\pacs{71.15.-m, 73.22.-f, 74.78.Fk, }
\maketitle



\section{\label{sec:level1}INTRODUCTION}
Two-dimensional layered materials are becoming the focus of experimental and theoretical investigations aiming to realize the potential applications of these atomically thin structures\cite{2dmaterial_rev1,2dmaterial_rev2,2dmaterial_rev3}. The library of these layered materials is still expanding with properties that range from metals and semi-metals to semiconductors and insulators.  Graphene, a semi-metallic atomically-thin sheet of carbon atoms in the honeycomb lattice\cite{graphene_2004,graphene_rev}, is an important member of the layered materials family. Flakes of graphene can be obtained by the exfoliation method from a graphite crystal\cite{graphene_exfoliation} or synthesized by methods such as chemical vapor deposition\cite{graphene_CVD}. In addition to its outstanding electronic and mechanical properties,  graphene is also an interesting platform to investigate the quasi-relativistic strongly-interacting many-body physics near the charge-neutrality point (CNP) when screening between charges is reduced\cite{graphen_dirac_fluid}. This electron-hole plasma, known as the Dirac fluid, behaves differently from the conventional Fermi liquid, when the Fermi level is far from the CNP. The inclusion of graphene in the van der Waals (vdW) heterostructures, that is, stacks of various layered materials, can serve several purposes, for instance as an active layer, a spacer or an electrode. The different layered materials exhibit a wide variety of physical properties such as topological phases\cite{TMDC_TI}, superconductivity\cite{TMDC_sc}, magnetism\cite{TMDC_magnetism} and charge density waves\cite{TMDC_cdw}. Different ways of stacking, manipulating these materials and intercalating with foreign atoms in the vdW heterostructure open even wider possibilities for interesting physics phenomena and for novel nano-device applications\cite{2dmaterial_rev2,layer_pseudospin}. Other control knobs also include electrical gating, an external magnetic field, various contacts and twist angles which affect strongly the Brillouin zone (BZ) alignment and the coupling between layers\cite{twist_graphene_review}. 

To investigate the electronic properties of these fascinating systems, large-scale density functional theory (DFT)\cite{graphene_sc_dft1,graphene_sc_dft2}, tight-binding models\cite{graphene_sc_tbh1,graphene_sc_tbh2,graphene_sc_tbh3,graphene_sc_tbh4}, and low-energy k $\cdot$ p expansions\cite{graphene_sc_kp1,graphene_sc_kp2,graphene_sc_kp3,graphene_sc_kp4} have been employed. The parameter-free DFT approach is computationally demanding, while the computationally efficient tight-binding methods and low-energy k $\cdot$ p expansions are hampered by the absence of a universal form of the interlayer couplings. The interlayer couplings employed by empirical methods are often parametrized as functions of only the interatomic distances or more elaborate forms that depend on the local bonding environment\cite{graphene_sc_tbh4,carbon_bond} with the values of parameters obtained by fitting the band structure of selected crystal configurations. A set of such interlayer hopping terms for graphene has been determined from {\it ab initio} calculations, but they are extracted only from a restricted {\em subset} of all possible bilayer orientations\cite{macdonald_graphene_bilayer,macdonald_moire}. The dependence of interlayer hopping on both the distance between pairs of atoms and the relative  orientation of bonds, as exemplified by the $\gamma_3$ and $\gamma_4$ terms in the Slonczewski-Weiss-McClure model\cite{graphene_rev}, have not yet been addressed properly. An accurate and transferable theory of interlayer coupling would not only provide an efficient way of evaluating electronic properties, but would also shed light on transport properties across layers\cite{moire_transport} and on the derivation of effective low-energy theories for arbitrary graphene stacking sequences.

We provide here a comprehensive and quantitive understanding of interlayer coupling in two prototypical bilayers, graphene-graphene (G-G) and graphene-hexagonal born nitride (G-hBN). For the first time, this type of {\it ab initio} modeling based on the Wannier transformation is applied to derive a transferable potential applicable to bilayer  configurations with arbitrary translations and twists between the two layers. In contrast to similar analysis applied to transition metal dichalcogenides, in which interlayer coupling was shown to have a simple, orientation-independent scaling form that depends only on the distance between pairs of atoms\cite{tmdc_tbh}, the G-G and G-hBN interlayer couplings include a dependence on both pair distances and relative orientations. The ensuing angular dependence is related to the crystal field distortions of the atomic $p_z$ Wannier orbitals. We derive the form of effective coupling terms that is suitable in the general vdW heterostructure with arbitrary twists and translations when some layers are integrated out. This scheme is relevant for obtaining the proximity effects on graphene due to other layered materials in the vdW heterostructure. These interlayer coupling models provide efficient ways for obtaining electronic properties and effective low-energy theories, as well as for estimating proximity effects in vdW heterostructures, especially when graphene layers are included in the stacks.

\begin{figure}
\centering
\includegraphics[width=0.5\textwidth]{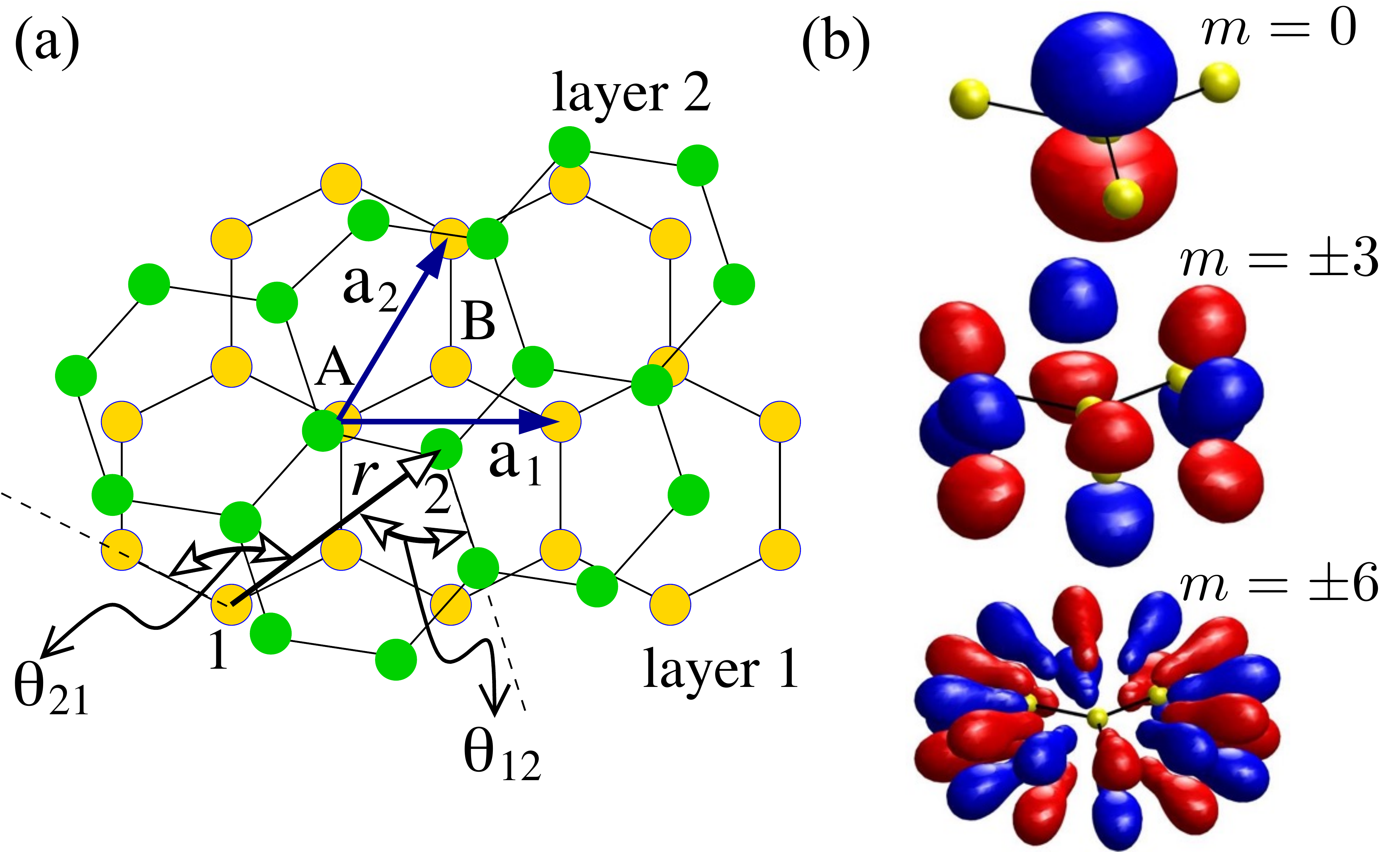}
\caption{(a) The generic graphene bilayer configuration with arbitrary translations and twists. The constituent monolayer crystal is described by the primitive vectors $\vec{a}_1$ and $\vec{a}_2$. For an interlayer pair, the coupling dependence is characterized by the projected distance $r$ and the angles $\theta_{12}$ and $\theta_{21}$ between $\vec{r}$ and the nearest neighbor bonds. (b) Decomposition of the Wannier function for monolayer graphene into the constant ($m=0$), $\cos(3\theta)$ ($m=\pm 3 $) and $\cos(6\theta)$ ($m=\pm 6 $) components.}
\label{fig:Wannier_orbital}
\end{figure}

The paper is structured as follows: In Sec. II, we introduce the numerical methods employed for the DFT calculations and the Wannier transformation. In Sec. III, we investigate the hamiltonians for weakly interacting bilayers, using G-G and G-hBN interfaces as the prototypical examples. In Sec. IV we discuss the physics of twisted bilayer graphene and in Sec. V we derive the effective theory for proximity effects by integrating out the neighboring layers. Our concluding Sec. VI gives a summary of the main points, makes comparisons with similar approaches in the literature, and contains some remarks on possible extensions and future applications.

\section{\label{sec:level1} NUMERICAL METHODS}
The approach we adopt here is to derive the {\it ab initio} tight binding hamiltonian based on the Wannier transformation of DFT calculations. Within DFT, we obtain the Bloch wavefunctions and energies using  VASP\cite{vasp1,vasp2} with pseudo-potentials of the Projector Augmented-Wave (PAW) type, the exchange-correlation functional of Perdew, Burke and Ernzerhof (PBE)\cite{pbe}, a plane-wave energy cutoff 500 eV and a 17 $\times$ 17 $\times$ 1 reciprocal space grid. A $20$ \AA $ $  distance is used to eliminate the coupling between periodic images of the layers in the direction perpendicular to the atomic planes. The diagonal Kohn-Sham hamiltonian in Bloch basis from the DFT calculations is then transformed into a basis of maximally-localized Wannier functions (MLWF)\cite{mlwf} implemented in the Wannier90 code.  In our modeling, only $p_z$-like orbitals at each atomic site are projected out and retained in the Wannier basis. The short-ranged {\it ab initio} tight-binding hamiltonian  we construct is an accurate and reliable way to obtain model parameters by preserving the phase and the orbital information from the DFT calculations. 

\section{\label{sec:level1} HAMILTONIAN FOR WEAKLY INTERACTING BILAYERS}
Before modeling the interlayer coupling, we first reconstruct the {\it ab-initio} tight-binding hamiltonian $\mathcal{H}_0$ for a graphene monolayer\cite{macdonald_graphene_mono}. The unit cell for monolayer graphene is spanned by $\vec{a}_1=(\sqrt{3}\hat{x}-\hat{y})a/2$ and $\vec{a}_2=(\sqrt{3}\hat{x}+\hat{y})a/2$ with the lattice constant $a$=2.46 \AA. Two basis atoms are situated at $\vec{\delta}_{\rm A}=0$ and $\vec{\delta}_{\rm B}=(\vec{a}_1+\vec{a}_2)/3$. We extract the intralayer couplings up to the eighth nearest neighbors, which shows good agreements with DFT results. The numerical parameters for $t_i$, the intralayer hopping parameter to the $i$-th nearest neighbor, are listed in the left block $\mathcal{H}_0$ of Table \ref{table:tbh_parm}.

\begin{table}[]
\centering
\caption{Graphene intralayer ($\mathcal{H}_0$) and interlayer ($\mathcal{H}'$ ) TBH parameters with the on-site energy $\epsilon_C=0.3504$ eV. $t_i$ (for the locations of these neighbors, see Ref. \cite{macdonald_graphene_mono}) and $\lambda_i$ are in eV with $a=2.46$ \AA $ $ for $\bar{r}$; $\xi_i$, $x_i$, $\kappa_i$ are dimensionless parameters.} 
\label{table:tbh_parm}
\begin{tabular}{cccccc|c|c|c|}
\cline{7-9}
\multicolumn{4}{c}{$\mathcal{H}_0$}                                                                                                          &                                  & $\mathcal{H}'$       & \textbf{$V_0(r)$} & \textbf{$V_3(r)$} & \textbf{$V_6(r)$} \\ \cline{1-4} \cline{6-9} 
\multicolumn{1}{|c|}{\textbf{$t_1$}} & \multicolumn{1}{c|}{$-2.8922$} & \multicolumn{1}{c|}{\textbf{$t_5$}} & \multicolumn{1}{c|}{0.0524}    & \multicolumn{1}{c|}{$\; \; \; $} & \textbf{$\lambda_i$} & $0.3155$          & $-0.0688$         & $-0.0083$         \\ \cline{1-4} \cline{6-9} 
\multicolumn{1}{|c|}{\textbf{$t_2$}} & \multicolumn{1}{c|}{0.2425}    & \multicolumn{1}{c|}{\textbf{$t_6$}} & \multicolumn{1}{c|}{$-0.0209$} & \multicolumn{1}{c|}{$\; \; \; $} & \textbf{$\xi_i$}     & $1.7543$          & $3.4692$          & $2.8764$          \\ \cline{1-4} \cline{6-9} 
\multicolumn{1}{|c|}{\textbf{$t_3$}} & \multicolumn{1}{c|}{$-0.2656$} & \multicolumn{1}{c|}{\textbf{$t_7$}} & \multicolumn{1}{c|}{$-0.0148$} & \multicolumn{1}{c|}{$\; \; \; $} & \textbf{$x_i$}       & $-$               & $0.5212$          & $1.5206$          \\ \cline{1-4} \cline{6-9} 
\multicolumn{1}{|c|}{\textbf{$t_4$}} & \multicolumn{1}{c|}{0.0235}    & \multicolumn{1}{c|}{\textbf{$t_8$}} & \multicolumn{1}{c|}{$-0.0211$} & \multicolumn{1}{c|}{$\; \; \; $} & \textbf{$\kappa_i$}  & $2.0010$          & $-$               & $1.5731$          \\ \cline{1-4} \cline{6-9} 
\end{tabular}
\end{table}

The shape of the localized basis, also known as the Wannier orbital provides intuition for the chemical bonding, hybridization and the symmetry of the crystal. For monolayer graphene, the constructed Wannier orbital has a dominant $p_z$ character but the azimuthal symmetry is broken by the crystal field distortion from the neighboring atoms. Locally, at the position of the carbon atom, the three-fold rotation symmetry is restored. Thus, the angular momentum is defined up to modulo 3, which means that there is hybridization within each sector of angular momentum states. In Fig. \ref{fig:Wannier_orbital}(b), we decompose the Wannier function for graphene into $m=0$ (dominant $p_z$), $m=\pm3$ and $m=\pm6$ angular momentum components. This decomposition shows the range and the strength of each component, and the characteristic radius gets larger for larger angular momentum components.

When two or more monolayers are brought into contact, the shape of the Wannier function has implications for the interlayer coupling. These couplings are described by the matrix elements: $\langle \psi_2|\mathcal{H}|\psi_1 \rangle$ with $\psi_1$ ($\psi_2$) the Wannier orbital of the first (second) layer and $\mathcal{H}$ the total hamiltonian. The angular momentum mixing as shown in Fig. \ref{fig:Wannier_orbital}(b) for the Wannier orbital in graphene translates into the angular dependence of such interlayer couplings, in addition to the usual dependence on the distance of the pair. Without loss of generality, we assume the projected vector $\vec{r}$ from $\psi_1$ to $\psi_2$ on the plane is along the positive $x$ axis, and $\theta_1$ ($\theta_2$) is the angle relative to $\vec{r}$ needed to determine the orientation of the crystal of the layer to which the Wannier orbital  $\psi_1$ ($\psi_2$) belongs. The interlayer coupling can then be written as the function $t(r,\theta_1,\theta_2)$. If the underlying crystal and the embedded Wannier orbital has N-fold rotation symmetry, then $\theta$ is only defined up to modulo $2\pi/N$. The above interlayer coupling can be simplified to:

\begin{equation}
t(r,\theta_1,\theta_2) = \sum_{m_1,m_2=-\infty} ^{\infty}f_{m_1,m_2}(r) e^{im_1 N_1 \theta_1+im_2 N_2 \theta_2}
\label{eqn:interlayer_general}
\end{equation} with integers $m_i$. For real $t$, $f_{\bar{m}_1,\bar{m}_2}(r)=f^*_{m_1,m_2}(r)$. This decomposition can be viewed as the multi-channel interlayer hopping process.

\begin{figure}
\centering
\includegraphics[width=0.5\textwidth]{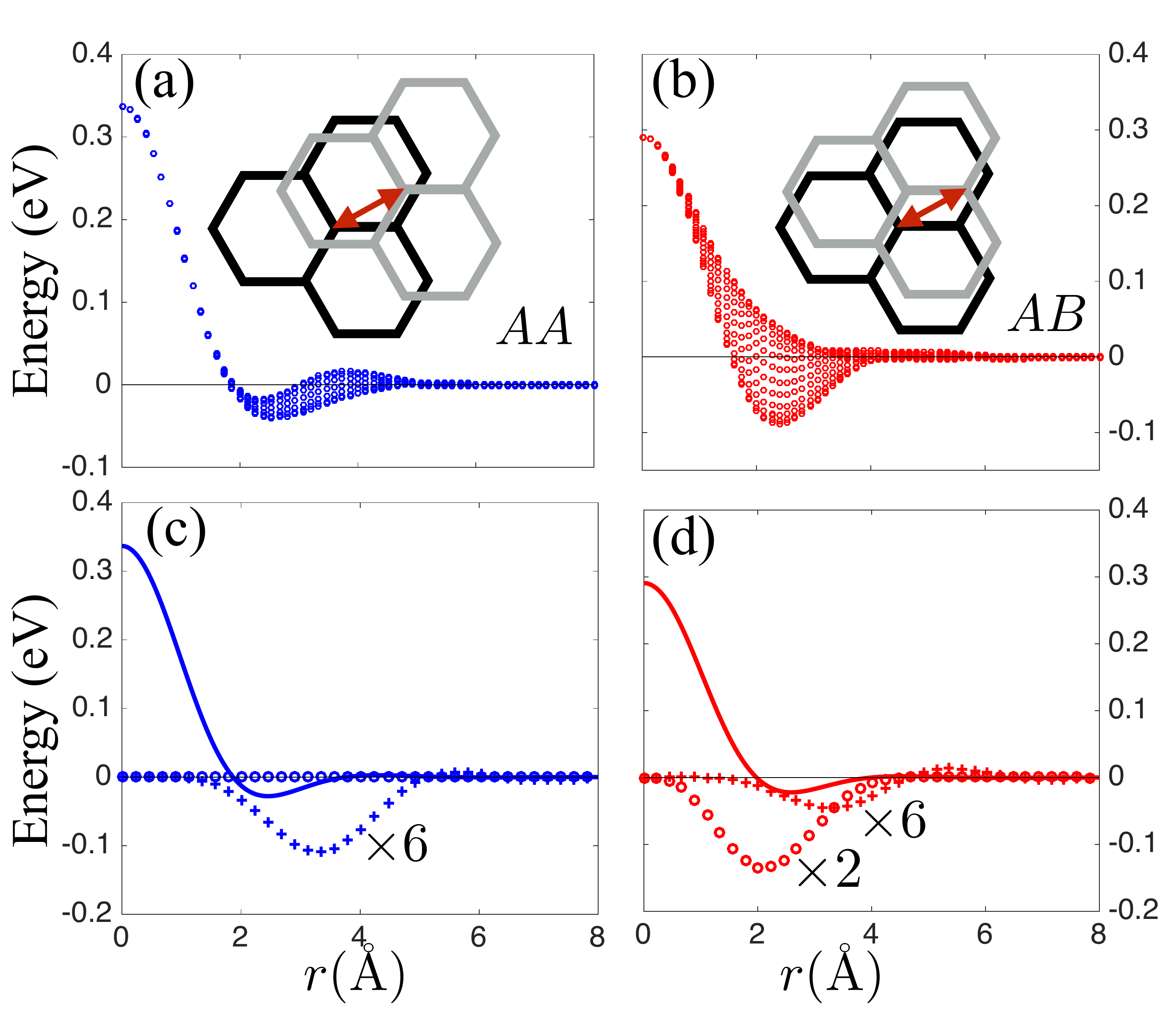}
\caption{Hoppings for the shifted graphene bilayer of (a) $AA$-type and (b) $AB$-type pairs as functions of the pair distance $r$; the spread of the hoppings at fixed $r$ indicates angular dependence. Decomposition into the constant $m=0$ (solid line), $\cos(3\theta)$ (circle), and $\cos(6\theta)$ (cross) components for the (c) $AA$-type pairs and (d) $AB$-type pairs. These curves are modeled by the $V_i(r)$ in Eq. (\ref{eqn:TBH_interlayer_fit}).}
\label{fig:GG_AA_AB}
\end{figure}


We next apply this general analysis to bilayers of graphene (G-G) and of graphene/hexagonal boron nitride (G-hBN). We consider two specific stackings, $AA$ and $AB$ defined by the relative position of the basis $A$ or $B$ atom of the top layer to that of the basis $A$ atom of the bottom layer. The bilayers are assumed to be flat with the same constant separation $c=3.35$ \AA $ $ in the $z$ direction\cite{graphene_rev}. Since in these two specific cases the two layers are not rotated with respect to each other, each primitive unit cell contains four atoms. After carrying out the DFT and Wannier transformation,, the {\it ab initio} tight binding hamiltonian $\mathcal{H}=\mathcal{H}_0^t+\mathcal{H}_0^b+\mathcal{H}'$ can be decomposed into the intralayer $\mathcal{H}_0^{t}$, $\mathcal{H}_0^{b}$ and the interlayer $\mathcal{H}'$ parts. The interlayer coupling of any atomic pair can be obtained from elements of $\mathcal{H}'$ in the Wannier basis. We then apply a lateral translation  $\vec{\Delta}=r\cos(\theta) \hat{x}+r\sin(\theta)\hat{y}$ to the top layer with the vertical separation $c$ fixed. This translation will affect both the distance and the relative orientation of the interlayer bonds while keeping the underlying crystal orientation untouched. The interlayer hoppings are extracted between the basis $A$ atom of the bottom layer at the origin and the shifted $A$ ($B$) basis atom of the top layer in the translated $AA$ ($AB$) structure.

The extracted interlayer hoppings as functions of the projected interlayer bond distance $r$ are plotted in Fig. \ref{fig:GG_AA_AB}(a), (b) for the graphene $AA$ and $AB$ bilayers, respectively. At a given distance $r$, the spread of the interlayer hopping indicates strong angular dependence. Notably, hopping in the $AA$ type bilayer is different from that in the $AB$ type bilayer. Due to the three-fold rotation symmetry of the underlying crystal, these interlayer hoppings are invariant under $\theta \rightarrow \theta \pm 2\pi/3$. We further decompose the angular dependence at fixed $r$ into its Fourier components of the constant term, $\cos(3\theta)$ and $\cos(6\theta)$ terms. Higher order $\cos(3N\theta)$ terms do exist but are vanishingly small. 

The hopping is given by the superposition of the interlayer terms that involve the symmetric combination of the following parameters

\begin{equation}
\begin{split}
t(\vec{r})= & V_0(r) +V_3(r) [\cos(3\theta_{\rm 12})+\cos(3\theta_{\rm 21})] \\
& + V_6(r) [\cos(6\theta_{\rm 12})+\cos(6\theta_{\rm 21})] 
\end{split}
\label{eqn:TBH_interlayer}
\end{equation} where $\vec{r}$ the two-dimensional (projected) vector connecting the two atoms, $r=|\vec{r}|$, and $\theta_{12}$ and $\theta_{21}$ the angles between the projected interlayer bond and the in-plane nearest neighbor bond as defined in Fig. \ref{fig:Wannier_orbital}(a). The result does not depend on which nearest neighbor bond is used. Compared with Eq. (\ref{eqn:interlayer_general}), these correspond to non-zero values for the terms $f_{0,0}$, $f_{\pm1,0}$, $f_{0,\pm1}$, $f_{\pm2,0}$ and $f_{0,\pm2}$, with $N_1=N_2=3$.   We use the following fitting functions for $V_i(r)$ with $\bar{r}=r/a$

\begin{equation}
\begin{split}
V_0(r) & =  \lambda_0 e^{-\xi_0 (\bar{r})^2} \cos(\kappa_0 \bar{r}) \\
V_3(r) & =  \lambda_3 \bar{r}^2 e^{-\xi_3(\bar{r}-x_3)^2} \\
V_6(r) & =  \lambda_6 e^{-\xi_6(\bar{r}-x_6)^2} \sin(\kappa_6 \bar{r})
\end{split}
\label{eqn:TBH_interlayer_fit}
\end{equation}

These Fourier projected components are plotted in Fig. \ref{fig:GG_AA_AB}(c) and (d) for the $AA$ and $AB$ stackings respectively. For the constant term, the $AA$/$AB$ hoppings are very similar to each other, and we define $V_0(r)$ to be the average of the two. Projection into $\cos(3\theta)$ is significant for the $AB$ type but vanishes identically for the $AA$ type, and the curve in the $AB$ case is defined as $2V_3(r)$. The two stackings have similar behavior for the much smaller $\cos(6\theta)$ term, and $2V_6(r)$ is the average of the two. The values of the fitting parameters are given in Table \ref{table:tbh_parm} from the analysis of the translated $AA$/$AB$ structures.

A proper analysis of the inherent symmetry of the two-layer system provides the justification for the form of the interlayer hoppings and enables us to generalize the model to arbitrary configuration. Specifically, the three-fold symmetry of the crystal field allows mixing between $p_z$ and $m=\pm 3N$ orbitals with $N$ an integer. Due to the crystal symmetry, these components acquire a phase $(-1)^N$ when $A$ and $B$ basis atoms are interchanged, which are related by a $yz$ mirror operation. Wannier orbital viewed as composite objects of mixed angular momentum, the hopping between two such objects is determined by the superposition of the individual hopping channels between each component, within the two center approximation\cite{slater}. The dominant channel is between the two $m=0$ components which gives the constant part in the interlayer hopping. There is a $\cos(3\theta)$ term from the coupling between $m=0$ and $m=\pm 3$ channels. Due to the symmetry, the two terms add up constructively (destructively) for the $AB$ ($AA$) type. This can also be seen from the additional minus sign in exchanging $A$ and $B$ basis atoms. There are two types of contribution to the $\cos(6\theta)$ term, and they can be generated from the coupling between $m=0$ and $m=\pm 6$, or between $m=\pm 3$ components of the two atoms. By symmetry, the first (second) part of the contribution is even (odd) in $AA$/$AB$. We can model the contribution of the first type from the average $\cos(6\theta)$ term of AA/AB in Fig. \ref{fig:GG_AA_AB}(c), (d). The channel between two $m=\pm 3$ can in general produce complicated angular dependence, but it is only a small correction (a few meV) and hence we ignore it in our model.

Following similar steps, we derive the tight-binding hamiltonian for interlayer coupling in the case of a bilayer G-hBN. In the G-hBN interlayer coupling compared to the G-G coupling the symmetry is lower since the atoms are not identical, and this affects the amplitude for each angular momentum channel. From similar analysis for the shifted $AA$/$AB$ couplings in a G-hBN bilayer, we can model the coupling as:

\begin{equation}
\begin{split}
t^{\rm CX}(\vec{r})=  &V^{\rm CX}_0(r) + V^{\rm CX}_{3, \rm XC}(r) \cos(3\theta_{\rm XC}) \\
& +V^{\rm CX}_{3,\rm CX}(r)  \cos(3\theta_{\rm CX})
\end{split}
\label{eqn:TBH_interlayer_GhBN}
\end{equation} where X=B,N atoms. $V^{\rm CX}_0(r) $, $V^{\rm CX}_{3, \rm CX}(r)$ and $V^{\rm CX}_{3, \rm XC}(r)$ share the same functional form as the ones in the G-G case, and the corresponding values of the parameters are tabulated in Table \ref{table:tbh_parm_hBN}. 


\begin{widetext}

\begin{table}[]
\centering
\caption{hBN intralayer and G-hBN interlayer tight-binding parameters.  $\epsilon_B=2.2021$ eV, $\epsilon_N=-1.9124$ eV with the convention $a=2.46$ \AA $ $ for $\bar{r}$ in the fitting expression Eq. (\ref{eqn:TBH_interlayer_fit}).}
\label{table:tbh_parm_hBN}
\begin{tabular}{cccccc|c|c|c|c|c|c|}
\cline{7-12}
\multicolumn{4}{c}{$\mathcal{H}_0^B$ ($\mathcal{H}_0^N$)}                                                                                  &                                  & $\mathcal{H}'$ & $V_0^{\rm CB}(r)$ & $V_{3,\rm BC}^{\rm CB}(r)$ & $V_{3,\rm CB}^{\rm CB}(r)$ & $V_0^{\rm CN}(r)$ & $V_{3, \rm NC}^{\rm CN}(r)$ & $V_{3,\rm CN}^{\rm CN}(r)$ \\ \cline{1-4} \cline{6-12} 
\multicolumn{1}{|c|}{$t_1$} & \multicolumn{1}{c|}{$-2.6490$}       & \multicolumn{1}{c|}{$t_5$} & \multicolumn{1}{c|}{0.0344 (0.0301)}     & \multicolumn{1}{c|}{$\; \; \; $} & $\lambda_i$    & 0.3905            & $-0.0588$                  & $-0.0651$                  & 0.2517            & $-0.0606$                   & $-0.0465$                  \\ \cline{1-4} \cline{6-12} 
\multicolumn{1}{|c|}{$t_2$} & \multicolumn{1}{c|}{0.0594 (0.2276)} & \multicolumn{1}{c|}{$t_6$} & \multicolumn{1}{c|}{$-0.0374 (-0.0240)$} & \multicolumn{1}{c|}{$\; \; \; $} & $\xi_i$        & 1.5426            & 3.0827                     & 3.7998                     & 1.6061            & 3.3502                      & 3.0464                     \\ \cline{1-4} \cline{6-12} 
\multicolumn{1}{|c|}{$t_3$} & \multicolumn{1}{c|}{$-0.2163$}       & \multicolumn{1}{c|}{$t_7$} & \multicolumn{1}{c|}{$-0.0053$}           & \multicolumn{1}{c|}{$\; \; \; $} & $x_i$          & $-$               & 0.6085                     & 0.6341                     & $-$               & 0.5142                      & 0.5264                     \\ \cline{1-4} \cline{6-12} 
\multicolumn{1}{|c|}{$t_4$} & \multicolumn{1}{c|}{0.0502}          & \multicolumn{1}{c|}{$t_8$} & \multicolumn{1}{c|}{$-0.0133$}           & \multicolumn{1}{c|}{$\; \; \; $} & $\kappa_i$     & 1.8229            & $-$                        & $-$                        & 2.1909            & $-$                         & $-$                        \\ \cline{1-4} \cline{6-12} 
\end{tabular}
\end{table}

\end{widetext}

\begin{figure}
\centering
\includegraphics[width=0.5\textwidth]{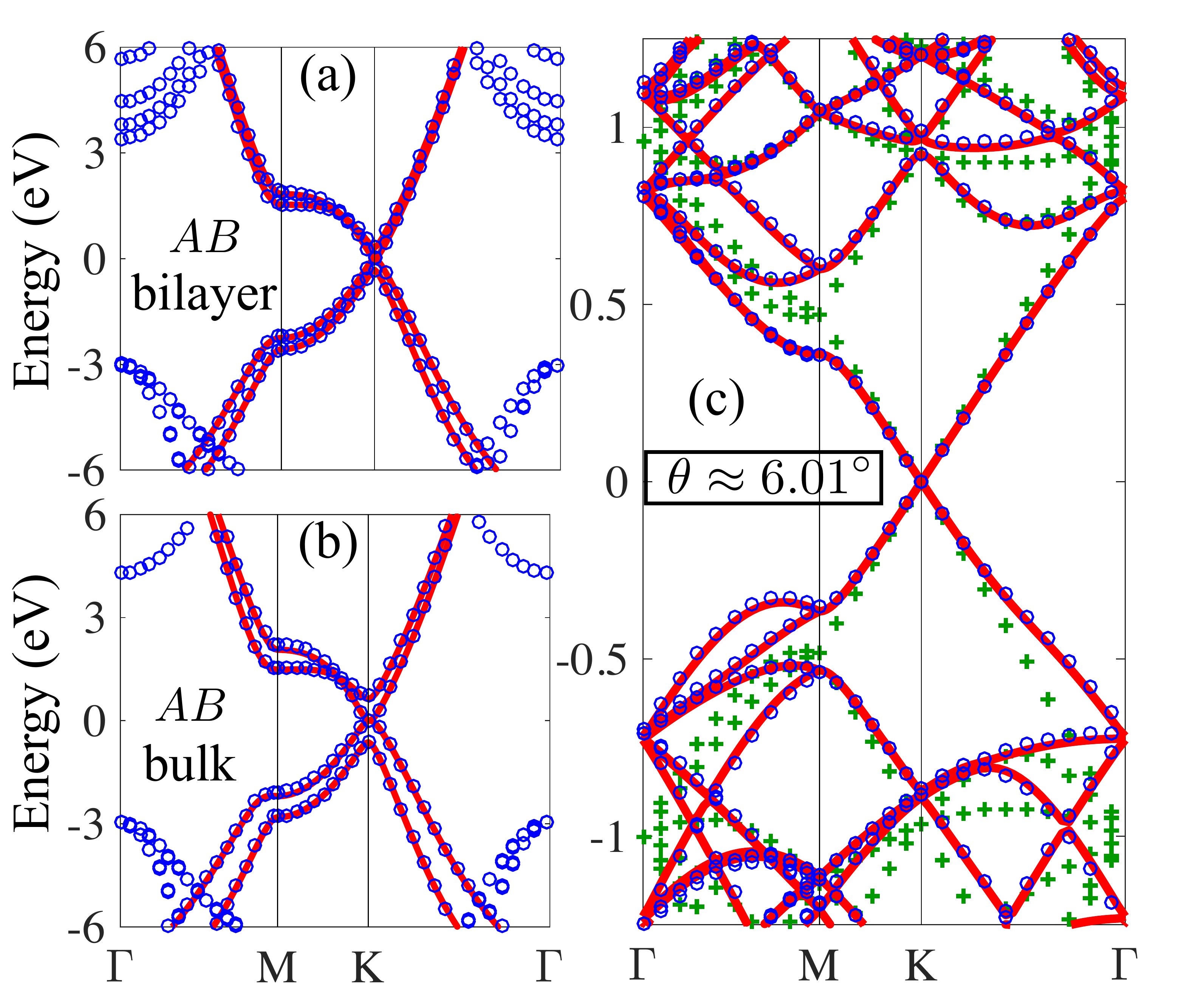}
\caption{Comparison between the tight binding hamiltonian (red lines) and {\it ab initio} DFT (blue circles) band structure calculations along $\Gamma$-M-K-$\Gamma$ in a $AB$ stacking (a) bilayer, (b) bulk; (c) $(M,N)$=(6,5) twisted super-structure ($\theta^{(6,5)} \approx 6.01^{\circ}$) and comparison to the folded monolayer band structure (green crosses).}
\label{fig:TBH_DFT_bands}
\end{figure}

To validate our model with the intra- and inter-layer couplings, in Fig. \ref{fig:TBH_DFT_bands}(a), (b) we compare the band structure obtained from DFT and from our tight-binding hamiltonian in the conventional $AB$-stacking bilayer and bulk graphite. The two band structures show good agreement over a large energy region around the Fermi level. The discrepancies away from the Fermi level are due to the hybridization of $p_z$ orbitals and other orbitals such as $sp_2$ which are not included in our Wannier model. When one monolayer is twisted relativ to the other, a supercell structure can be constructed in the commensurate case\cite{graphene_sc_dft1,SC_def}, labeled by ($M$,$N$) with twist angle $\theta^{(M,N)}$. The two layers are separated by a constant height $c=3.35$ \AA. In Fig. \ref{fig:TBH_DFT_bands}(c), we compare the result from DFT and tight-binding calculations for the ($M$,$N$)=$(6,5)$ twisted super-cell ($\theta\approx 6.01^{\circ}$); the model hamiltonian reproduces the DFT band structure well. We also include in Fig. \ref{fig:TBH_DFT_bands}(c) a comparison with the band structure of the folded BZ for a single monolayer, which is quite different, showing the importance of having an accurate description of interlayer coupling.

\section{\label{sec:level1} TWISTED BILAYER GRAPHENE PHYSICS}

\begin{figure}
\centering
\includegraphics[width=0.5\textwidth]{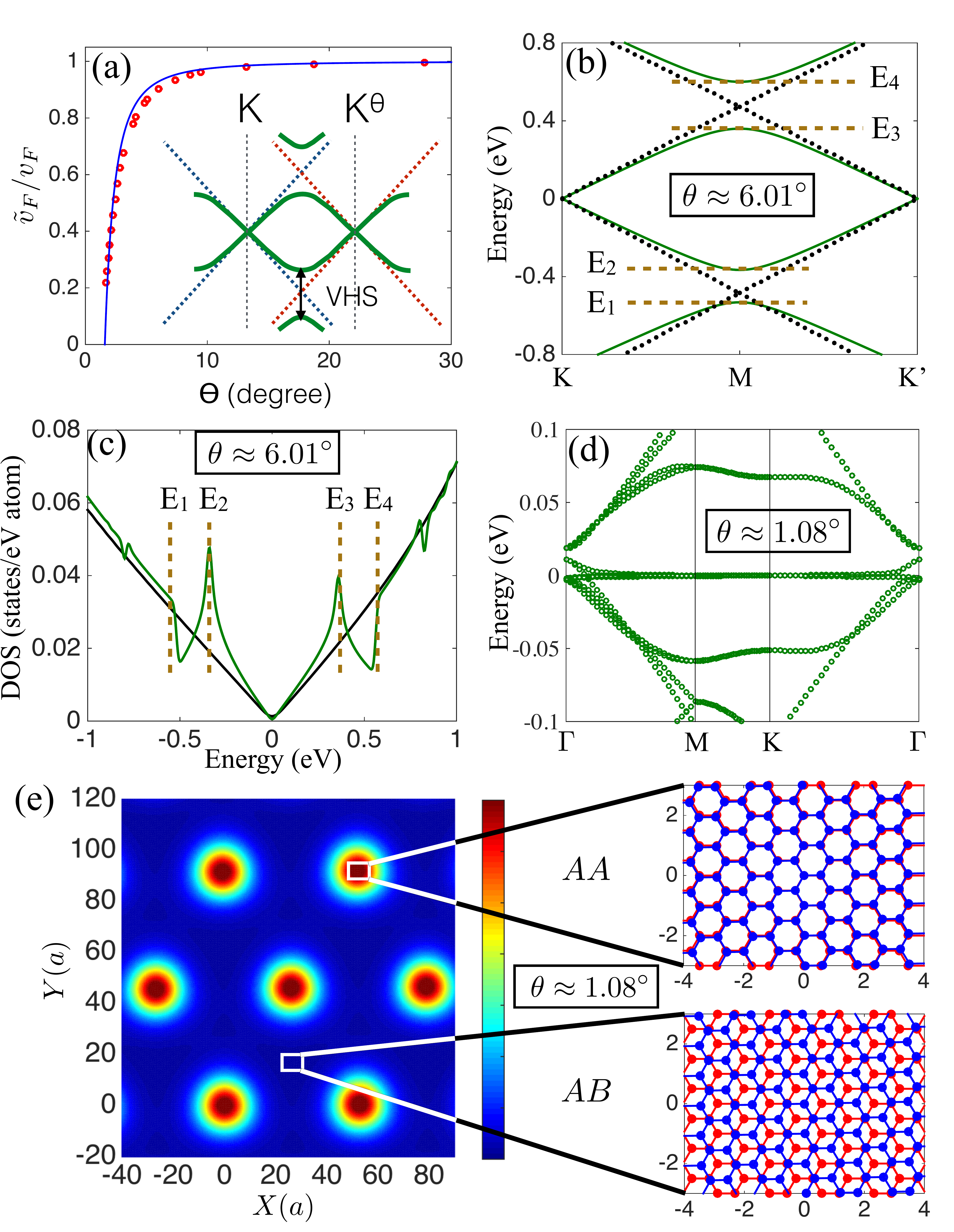}
\caption{(a) The renormalized Fermi velocity, $\tilde{v}_F/v_F$, as a function of the twist angle $\theta$ and VHS in the DOS (inset). (b) The band structure along K-M-K' in a $(M,N)$=(6,5) supercell ($\theta \approx 6.01^\circ$) compared to the folded monolayer bands (black dots). (c) The DOS for a $(M,N)$=(6,5) supercell (green) compared with the DOS of the monolayer (black). The energy extrema in the bands in (b) correspond to the singular points of DOS. (d) Similar calculations for the $(M,N)$=(31,30) supercell ($\theta\approx1.08^\circ$). (e) The states around the Fermi level are localized at $AA$-sites forming a Moir\'{e} pattern.}
\label{fig:twist_physics}
\end{figure}

The coupling between layers is weak for $\theta \approx 30^\circ$, and gets stronger when the twist approaches angles near $0^\circ$ or $60^\circ$.\cite{graphene_sc_tbh1} When two layers are twisted, the bands are formed from the hybridization of monolayer bands\cite{graphene_sc_tbh3} as in the schematic diagram of Fig. \ref{fig:twist_physics}(a) inset. The characteristic kinetic energy scale is defined by $\hbar v_F \Delta K$, with $v_F=8.22 \times 10^5$ m/s the Fermi velocity of the monolayer graphene, and $\Delta {\rm K} =|{\rm K}-{\rm K}^\theta| = \frac{8\pi}{3a} \sin(\theta/2)$ the distance between displaced K points of two layers (${\rm K}=\frac{4\pi}{3a} \hat{y}$, ${\rm K}^\theta=\frac{4\pi}{3a} (\cos(\theta)\hat{y}-\sin(\theta)\hat{x})$). Under the hybridization at large twist angles, the bilayer bands retain the linear Dirac dispersion, but with a different slope around the Dirac point compared to the folded bands of the monolayer graphene as in Fig. \ref{fig:TBH_DFT_bands}(c) and \ref{fig:twist_physics}(b). The Fermi velocity is renormalized by the interlayer coupling as can be seen from the effective low-energy theory around the Dirac points. A pair of states around the Dirac point with near zero energy are coupled to three pairs of states of energy $\pm \hbar v_F \Delta K$. The effective theory from a Schrieffer-Wolff transformation for the near zero energy doublet states has corrections to linear order in $k$ which renormalize the velocity\cite{graphene_sc_kp1}. We constructed a series of twisted supercell structures with decreasing angles from $30^\circ$, and the renormalized Fermi velocity calculated from the bands along $\Gamma$-K indeed follows the theoretical prediction $\tilde{v}_F /v_F = 1-C /\sin^{2} (\theta/2)$ with $C = 1.953 \times 10^{-4}$.\cite{graphene_sc_kp1}

Another feature for the band structure of the twisted superstructure is the van Hove singularities (VHS) in the density of states (DOS) near the Fermi level\cite{graphene_sc_tbh3}, which often leads to electronic instabilities such as superconductivity\cite{vanhove_sc} and magnetism\cite{vanhove_mag} in the many-body system. In Fig. \ref{fig:twist_physics}(c), the DOS of a $(M,N)$=(6,5) supercell with $\theta \approx 6.01^\circ$ is compared to a monolayer graphene with the singular points corresponding to the energy extrema in Fig. \ref{fig:twist_physics}(b). This VHS is due to gap opening from hybridization between states in the overlap between the Dirac cones of the two layers \cite{graphene_sc_tbh3}. The advantage of twisted bilayers is that the location of VHS can be controlled by varying the twist angle\cite{van_hove}, and are roughly centered at $E=\pm \frac{1}{2}\hbar v_F \Delta K$. 

When the twist angle is even smaller, such as with a $(M,N)$=(31,30) supercell structure ($\theta \approx 1.08^\circ$), the Fermi velocity is close to zero, and nearly flat bands are observed at the Fermi level\cite{graphene_sc_tbh2} in Fig. \ref{fig:twist_physics}(d). The electronic states in these nearly dispersionless bands show highly localized charge density at the $AA$-sites\cite{graphene_sc_tbh1} as in Fig. \ref{fig:twist_physics}(e), referred to as Moir\'{e} pattern. In experiments, the VHS and the localization of electrons in twisted graphene layers have been measured by scanning tunneling spectroscopy\cite{van_hove,moire_exp,moire_exp2}.

In the discussion so far we have assumed the layers to be flat and focused only on their electronic properties. Structural relaxations such as rippling or more drastic commensurate-incommensurate transitions with domain-line formation, as in the G-hBN bilayer\cite{ghbn_com_incom}, could be relevant for small twist angles. They are driven by the different local mechanical energy for $AA$ and $AB$ stackings\cite{graphene_com_incom}. Though the prediction of mechanical deformations is beyond the scope of the current work, we comment that more general forms of interlayer couplings can be modeled by incorporating variable height or strain by modifying the initial $AA$/$AB$ sliding bilayers, which will allow proper description of the effects of structural deformations.

\section{\label{sec:level1} EFFECTIVE THEORY FROM PROXIMITY EFFECTS}
As a final comment on how our model can be applied, we discuss how to construct effective theories with limited degrees of freedom instead of having to solve the hamiltonian in the full Hilbert space of the combined layers. In graphene bilayers, the low-energy hamiltonian has the form of non-abelian gauge theory\cite{non_abelian_gauge}. To define and formulate the problem, we consider a vdW bilayer heterostructure with layer 1 as the main component where the low-energy degrees of freedom at the Fermi level $E_{F}$ reside. Layer 2 is brought close to layer 1 to introduce the desired proximity effects. In general, the presence of layer 2 will affect the electronic properties of layer 1 in two ways: (1) by introducing a direct additional potential generated by the neighboring atoms; (2) by introducing virtual interlayer hopping processes through hybridization to the states of the neighboring layer. The general vdW heterostructure hamiltonian takes the form

\begin{equation}
\mathcal{H}_{\rm vdW}=\begin{bmatrix}
\hat{H}_{1} + \Delta \hat{H}_{12}  & \hat{T} \\
\hat{T}^\dagger & \hat{H}_{2} + \Delta \hat{H}_{21}
\end{bmatrix}
\end{equation} $\Delta \hat{H}_{12}$ and  $\Delta \hat{H}_{21}$ are the direct corrections of the first type, and $\hat{T}$ is the interlayer coupling in the heterostructure. Integrating out the second layer gives a perturbation term for the first layer

\begin{equation}
\Delta V_{12}(E)=\hat{T} \frac{1}{E-(\hat{H}_2+\Delta \hat{H}_{21})} \hat{T}^\dagger
\label{eqn:eff_layer1}
\end{equation} where, since $\hat{T}$ is already small, and $\Delta \hat{H}_{21}$ is a higher order correction that disrupts the lattice translation symmetry in $\hat{H}_2$, this term can be ignored. The effective potential takes the following form in the spatial representation, evaluated at $E=E_F$

\begin{equation}
\begin{split}
& \Delta V_{12}(r_2,r_1,E_F) = \sum_{s_2,s_1}  t_{r_2,s_2}   \langle s_2| \frac{1}{E_F-\hat{H}_2} |s_1 \rangle  t^\dagger_{s_1,r_1} \\
&=\sum_{s_2,s_1}  \frac{t_{r_2,s_2} t^\dagger_{s_1,r_1}  }{\Omega_k} \int_{\rm BZ} d^2 \vec{k}  \langle \psi_{\vec{k},\delta_2}| \frac{ e^{i \vec{k}\cdot (\vec{s}_{2}-\vec{s}_1)}  }{E_F-\hat{H}_2(\vec{k})}|\psi_{ \vec{k},\delta_1} \rangle
\end{split}
\end{equation} with $\Omega_k$ the BZ area, $r_i$ ($s_i$) the localized orbitals of layer 1 (2), which include both the position vector $\vec{r}_{i}$ ($\vec{s}_i$) and the orbital index $\delta_i$, and $t_{i,j}$ are the interlayer coupling from $j$ to $i$ orbitals between the layers. In the usual perturbation framework, this expression describes hopping across the layers from $r_1$ to $s_1$, allowing for all paths $s_1$ to $s_2$ within layer 2, and hopping back to layer 1, from $s_2$ to $r_2$. When applied to the G-hBN bilayer, the sub-lattice symmetry breaking mass terms of $\Delta V_{12}$ from hBN to carbon sites have opposite sign from the direct term $\Delta H_{12}$: the carbon site above a BN layer experiences the same sub-lattice potential as the BN layer itself in the direct contribution $\Delta \hat{H}_{12}$ while $\Delta V_{12}$ is opposite due to level repulsion in the framework of perturbation theory. The use of this effective potential will enable application to very large systems without loss of accuracy.

\section{\label{sec:level1} CONCLUSION}

In summary, we derived the {\it ab initio} G-G and G-hBN interlayer couplings based on the maximally localized Wannier function transformation of DFT calculations. We show that these interlayer couplings have both pair-distance and angular-orientation dependence. In contrast, the conventional way of modeling such couplings by fitting band structure calculations leads to ambiguities in the functional form and its dependence on important structural variables\cite{graphene_sc_tbh1,graphene_sc_tbh2,graphene_sc_tbh3,graphene_sc_tbh4, carbon_bond,fit_tb}. The success of the latter, simpler approach is due to the small number of parameters needed in effective low-energy theories near the Dirac energy\cite{graphene_sc_kp1,graphene_sc_kp2,graphene_sc_kp3,graphene_sc_kp4} implying that only one set of dominant Fourier components for interlayer coupling is relevant; this set of components, however,  is not enough to constrain its functional form and its dependence on key variables. In the work by Jeil Jung {\it et al.}  \cite{macdonald_graphene_bilayer,macdonald_moire}, such interlayer couplings were extracted with the use of the Wannier transformation but the crystal configuration of the bilayer was held at fixed orientation which means that it can only be applied to layered stacks that involve only translations and small relative twist angles.

In our work, we elucidate the physics from the extracted couplings by analyzing the multi-angular-momentum channel contributions.  This enabled us to generalize the interlayer coupling model to arbitrary stacking orientations with that involve any possible relative translation or rotation of the layers. Our model can also be generalized to incorporate local variations of in-plane strain and interlayer distance by varying the reference configurations in a systematic way. We expect our model to be relevant in investigating the derivation of low-energy theories appropriate for layer stackings\cite{graphene_sc_kp1,graphene_sc_kp2,graphene_sc_kp3,graphene_sc_kp4,non_abelian_gauge}, the band gap introduced by the presence of hBN\cite{mac_donald_gap_ghbn}, optical absorption\cite{optical_absorp}, vertical transport across layers\cite{moire_transport}, phenomena like the quantum Hall effects and Hofstadter's butterfly from the competition between magnetic field and supercell length scales\cite{QHE_TwBLG,moire_butterfly,kim_butterfly}. The systematic Wannier approach also allows for further generalization to other two-dimensional layered materials.

\begin{acknowledgements}
We thank Dennis Huang, Daniel Massatt, Stephen Carr, Min-Feng Tu, Sharmila N. Shirodkar, Georgios A. Tritsaris, Paul Cazeaux, Eric Cances, Mitchell Luskin, Bertrand Halperin and Philip Kim for useful discussions. This work was supported by the STC Center for Integrated Quantum Materials, NSF Grant No. DMR-1231319 and by ARO MURI Award W911NF-14-0247. We used the Extreme Science and Engineering Discovery Environment (XSEDE), which is supported by NSF Grant No. ACI-1053575.
\end{acknowledgements}



\begin{thebibliography}{1}
\bibitem{SC_def} The $(M,N)$ supercell has twist angle $\cos (\theta^{ (M,N)}) = \frac{N^2+4NM+M^2}{2(N^2+NM+M^2)}$. The supercell of the bottom layer has cell vectors $\vec{a}^{(M,N)}_1 = N\vec{a}_1 + M\vec{a}_2$ and $\vec{a}^{(M,N)}_2 = -M\vec{a}_1 +(M+N)\vec{a}_2$ while the top layer has cell vectors $\vec{a'}^{(M,N)}_1 = M\vec{a'}_1 + N\vec{a'}_2$ and $\vec{a'}^{(M,N)}_2 = -N\vec{a'}_1 +(M+N)\vec{a'}_2$ with $\vec{a'}_i$, the counterclockwise-rotation of $\vec{a}_i$ by $\theta^{(M,N)}$.
\end{thebibliography}
\end{document}